\documentclass[prb,preprint]{revtex4-1} 

\usepackage{amsmath}  
\usepackage{amsfonts} 
\usepackage{graphicx} 
\usepackage{xcolor}
\usepackage{csquotes}
\usepackage{hyperref}
\usepackage{braket}
\usepackage{mathtools}
\usepackage{comment}

\begin{document}


\title{A unified picture of Balance puzzles and Group testing : Some lessons from quantum mechanics for the pandemic}

\author{Chetan Waghela}
\email{chetan.waghela@iitrpr.ac.in} 
\affiliation{Department of Physics, Indian Institute of Technology Ropar, Rupnagar, Punjab-140001.}



\date{\today}

\begin{abstract}

Balance (Counterfeit coin) puzzles have been part of recreational mathematics for a few decades. A particular type of Counterfeit coin puzzle is known in the literature as the "Beam balance puzzle". An abstract solution to it is provided by Iwama et.al as a modification of the Bernstein-Vazirani algorithm, making use of quantum parallelism and entanglement. Moreover, during this pandemic, group testing has proved to be an efficient algorithm to save time and cost of testing specimens for the presence of infection. In this article, we propose a "Binary Spring Balance" (BSB) puzzle, to facilitate a unified picture of the counterfeit coin problem and the testing for infection problem, as both aim to reduce the number of queries. We then showcase two solutions to the BSB problem, one using bits and other using classical-qubits ('cebits") for querying. Both solutions are demonstrated using circuits. In this pursuit, we develop a modified optical implementation of Bernstein-Vazirani algorithm using only polarizers (no need of beam splitters), which has surprisingly not yet been proposed earlier. Under the pretext of this demonstration we question why we have not yet developed testing mechanisms inspired by Bernstein-Vazirani algorithm for the pandemic, as they solve the problem in single query, they have no issues related to prevalence of infection in the population, nor are they plagued by the issue of dilution of samples due to pooling. The modified implementation of Bernstein-Vazirani algorithm using polarizers can also be a cost-effective demonstration in an undergraduate lab.   
\end{abstract}

\maketitle 

\section{Introduction} 

Testing (or querying) is a process to reveal or enhance a particular hidden feature of a specimen. For example, for a counterfeit coin of a higher weight among original coins which are identical on observation, the only way to reveal the counterfeit coin is to weigh them. Similar to the case with coins, all identical looking swabs can be identified for the presence of infection using an RT-PCR test (there are other tests also available but for mathematical abstraction they are all considered equivalent).

A simplified description of the commonly used RT-PCR tests can be found at \cite{RTPCR}. The exact details of the testing are not required for the purpose of this article.

An uninformed person would test or weigh each of the swabs or coins individually. However, in the event when there are too many specimens to be tested it is tedious and in some cases impossible to test all of them individually. The time and testing kit cost shoots in proportion to the sample size. An innovative technique had been invented to counter this, which is known as "Group testing" or "Pooled testing", to identify the positive specimen in least number of tests \cite{Naturemath}. Similarly, the problem to identify counterfeit coins in least number of weighing is known as the "Balance Puzzle" or the "Counterfeit coin puzzle" \cite{TanyaK}. 

The testing mechanism depends on the nature of the sample. Balance puzzles use weighing as testing mechanism while identification for infection uses various laboratory methods. Moreover, the instruments used for weighing or identification for infection can differ and cause a change in mathematical structure of the solution (or algorithm) to the problem. For example, there are two types of instruments that can be used to weigh for the Balance puzzles. One being Beam balance while the other being Spring balance. The beam balance works by comparing mass. Equal number of coins (whose total will always be even) are kept on either side and then compared, if the pan tilts then a counterfeit coin is present in either of the pan. The knowledge of the weight of the coin is not needed for it. On the other hand spring balance does not function by comparison of weight. If a spring balance is used, knowledge of the weight of original coin is at least necessary. The weight shown on the scale is compared to the predicted weight depending on the number of coins placed. Then, if it is different compared to the predicted weight (assuming all coins are original), a counterfeit coin is present in the weighed coins. Additionally, spring balance also reveals the number of counterfeit coins in a particular weighing (deduced from the deficit between the value shown by the weighing and predicted value if all are considered original). 

We will study a unified logical picture of both the problems of spring balance and testing for infection in a population, and call it "Binary spring balance" (BSB) puzzle. It is to be noted that the testing mechanism for identifying positive specimens is similar to that of the application of spring balance to find counterfeit coins, and it is not similar to the application of beam balance. This is because a beam balance works by comparison of coins and always needs even number of coins for a particular weighing \cite{Iwama}. In the BSB puzzle the samples are bits rather than coins of swabs (0 denoting original bit and 1 denoting defective bit). The goal of this puzzle is the same i.e. to identify the defective bits using least number of queries. This puzzle demonstrates a unified logical picture of both the problems.

We also intend to showcase two circuits to implement on this binary sample. One where queries cannot be superimposed while in other case where queries can be superimposed. The second circuit is inspired by the Bernstein-Vazirani algorithm, it however works in the classical domain, too. The second circuit may pave a path for creating a framework to develop testing mechanism and algorithms which work by using only a single test for population of any size.

The article is arranged as follows: In Section 2, we will discuss the history and literature of problems individually. In Section 3, we discuss Li's S stage algorithm implemented for identification of positive specimens in a population. In Section 4, we will introduce the "Binary Spring Balance" puzzle and showcase a circuit which implement's Li's S stage strategy (where queries are not superimposed) for the problem with binary string size N=12. Section 5, we will showcase a circuit which solves the same problem using Bernstein-Vazirani algorithm (where queries are superimposed). In the same section we showcase a circuit to implement the algorithm using just polarizers (linear polarizers and half wave retarders). In Section 6, we will finally conclude. 

\section{History of the problems}
\subsection{Pooled or group testing}
The earliest mention of the pooled testing strategy was in the year 1943 \cite{Dorfman, Zhu}. In 1943, during the Second World War the United States Public Health Service took up a task to remove all syphilitic men called up for induction. testing required drawing a blood sample from an individual and then analysing the sample to determine the presence or absence of syphilis. During those times, performing this test was expensive due to less resources. Testing every soldier individually would have been very expensive and also would need lot of time.  This is because, if a person is not syphilitic then the testing kit as well as time is wasted on testing the person. Hence, in the scenario when the number of tests are large and the number of infected (or defective or counterfeit) samples are comparatively less, individual testing is an inefficient method as we will see.  

\begin{figure}
    \centering
    \includegraphics[width=1.0\columnwidth]{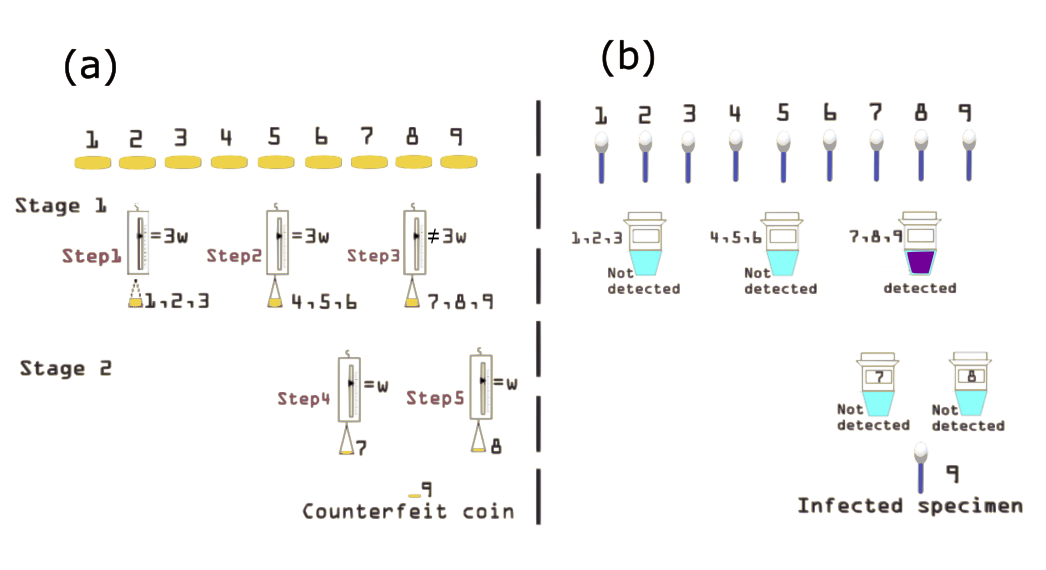}
    \caption{Schematic comparison of Spring Balance puzzle and Dorfman's (or Li's S=2 stage) solution for pooled testing for identifying positive specimens. (a) Example with N=9 coins with 9$^{th}$ coin being counterfeit. (b) Example with N=9 swabs with 9$^{th}$ swab being positive for infection. Both examples have been solved in 2 stages using Li's S stage algorithm and groups of size k=3 in 1$^{st}$ stage. In both solutions the samples are pooled together and tested. In the case of coins if the weight is not equal to 3 times the weight of each original coin (w), then a counterfeit coin is present in the group. In the case of swabs, if the color changes to dark blue in comparison to light blue for negative pools, then an positive specimen is present in the pool. A total of 5 queries are needed to solve both of these particular examples.}
    \label{balancedorfman}
\end{figure}

Dorfman proposed a clever way to solve this problem by pooling the blood samples.  A graphical depiction of Dorfman testing is depicted in Fig. \ref{balancedorfman}(b). If there are 100 men to be tested then samples can be grouped in batches of 10. The samples can then be tested in group. If any of the group shows sign of syphilis then each sample in the group can be tested individually. In this way, we can reduce the number of tests. There are 2 stages in this scheme, first is testing the complete population in groups and then individually testing the identified infected groups. This strategy is not the most efficient and it can be improved upon as shown by later literature. There are various modifications of this problem for various scenarios. In 1957, Sterrett \cite{Sterrett} proposed a modification of Dorfman's procedure. In 1958, Sobel and Groll \cite{Sobel} improvised the solutions for various scenarios and for the first time studied this problem in the context of Information theory. The reader is advised to follow \cite{Zhu} for more information related to the history of the research on group testing.

A variant of the pooling strategy is known as "Matrix pooling" or "Array querying" \cite{Naturemath}. In these solutions a sample is shared between two or more pools in same stage rather than only one pool. Recently, a method using Kirkman Triples, known as "Tapestry pooling" has been proposed by \cite{Tapestry}. In this strategy too the samples are distributed in different groups in same stage. It is proposed that in this scheme only one stage of querying is needed to identify all infected. However, there are issues like dilution of samples due to multiple distribution and also in keeping track of the strategy, which needs additional solutions.

\begin{figure}
    \centering
    \includegraphics[width=0.8\columnwidth]{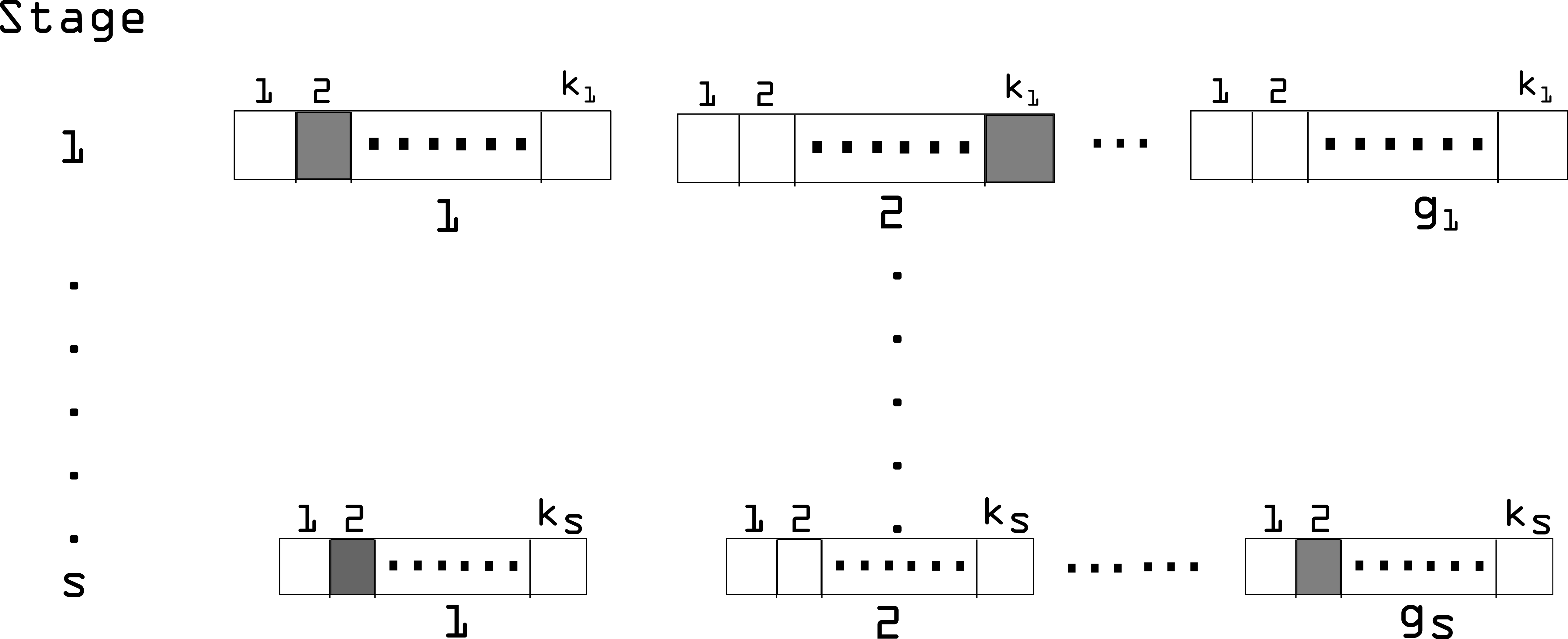}
    \caption{Schematic representation of Li's S stage algorithm for finding d positive specimens in a population of size N. $k_i$ is pool size at i$^{th}$ stage and the number of groups formed at i$^{th}$ stage are $g_i$. According to this nomenclature, $k_1 \times g_1=N$. Grey boxes represent positive specimens for pictorial purposes. Dorfman's strategy only has S=2 stages.}
    \label{Lis}
\end{figure}

In 1962, Li\cite{Li} came up with a generalization of the Dorfman's 2 stage procedure, called as S stage procedure Fig. \ref{Lis}. The difference in Li's S stage procedure is to regroup the detected groups with infection in second stage and then retest, repeating this procedure for 'S' number of stages rather than just 2 as in the  case of Dorfman's scheme. This procedure is efficient compared to the Dorfman's procedure as it further reduces use of testing kits as well as time. This procedure is also easier to implement and reduces human error compared to other schemes like Matrix pooling or Tapestry pooling. 

We will study pooling strategies in terms of Li's generalized procedure for our paper due to its' simplicity, it being generalization of the Dorfman's procedure which has received widespread implementation during recent Covid-19 pandemic \cite{Naturemath, Lagopati}.

Some important parameters to consider are the accuracy of the tests (i.e. sensitivity and specificity of the tests \cite{Swift}, the "prevalence rate" and the "pool size"\cite{Zhu}. For this paper we will assume that the tests are always accurate. The prevalence rate is the ratio of the number of infected to the total population to be tested. Pool size is the number of the samples to be included in one pool. Prevalence rate plays an important role in determining optimal pool size for pooling and also to determine if the pooling strategy is any better than individual testing. We also assume that the maximum number of specimens that can pooled together for efficient detection is unlimited (this is not true in reality due to  dilution). This is quantitatively discussed later in detail. 

We will explicitly define the specific problem that we will consider for the study:

\begin{displayquote}
You have N similar specimens from suspected patients and a test for the presence of the infection. All the specimens are indistinguishable without the test. 'd' number of specimens are infected and show unique marker compared to non-infected ones upon testing. What is the maximum number of tests required by the Li's pooling strategy considering 'g' pools of size 'k'.
\end{displayquote}

We are interested in the maximum number of tests required because it prepares us for the worst case scenario. The worst case scenario for a pooling strategy like Li's occur when the 'd' positive specimens are evenly distributed among 'g' pools of size 'k' made from a population of size 'N'. Suppose for 18 specimens, 2 are infected (this number is unknown to you in reality). They are each distributed in separate 2 of the 3 pools of size of size 6, each. Application of the pooling strategy would require maximum of 13 tests combined (3 in first stage and at most 10 in second). If the infected were all in same pool, it would require 9 tests only (3 in first stage and maximum of 6 in second stage of individual testing). As we cannot know before hand if the infected samples are evenly distributed or not, we need to be prepared for 13 tests in total, known as worst case scenario.

Suppose for 100 positive specimens 5 are infected ones, and there is one positive specimen in each of the groups. Using Dorfman's strategy on 2 stages, it would require testing each and every group individually and this increases the number of total tests to 110 (a number higher than the actual population size of 100). On the other hand, if the 10 positive specimens were in a same group, it would reduce the number of maximum tests required to only 10. We cannot control if the specimens would be evenly distributed or not, hence we should consider worst case scenario.

\subsection{Balance Puzzle or the Counterfeit coin puzzle}
The earliest known mention of the balance puzzle (or also known as the counterfeit coin problem) is from the year 1945. The problem was proposed by E.D. Schell in the American Mathematical Monthly :\cite{Schell}

\begin{displayquote}
You have eight similar coins and a beam balance. At most one coin is counterfeit and hence underweight. How can you detect if there is an underweight coin, and if so then which one, using the balance only twice?
\end{displayquote}

Over the years, there have been many generalizations of this problem, like \cite{Bonis, Pyber, Hwang, Bundy, Linial}. Various improvisations over the solution have also been proposed.

The testing mechanism proposed in these problems work by placing equal number of coins on both pans of a beam balance and then observing for  a tilt in the pan height. If the balance is tilted on either side, it shows that counterfeit coin is present in either of them. 

An uninformed person would place single coin on each of the pans of the beam balance and then try to observe if there is any tilt and then repeat the process until the counterfeit coins are identified. This would be a tedious process if the number of coins are large and number of counterfeit coins are comparatively very less. The method to solve this problem in lesser number of weighing in such scenario is by grouping (or pooling) the coins and then weighing. This reduces the number of weighing (tests) required if the number of counterfeit coins are very less.

Usually, for the standard balance puzzle, a beam balance is used. In this paper we will consider the balance puzzle with a spring balance. Fig. \ref{balancedorfman}(a) depicts the algorithm to detect a single counterfeit coin among 9 identical coins using spring balance. Each original coins weight=w. When the coins are pooled together and weighed they should weigh 3w if all are original. If there is any deficit then there is a counterfeit coin present. The reason to consider this is that in the standard beam balance puzzle only even number of coins can be tested at once, however, for the spring balance any number of coins can be tested at once. The testing mechanism is slightly different from application of beam balance above. The details are as discussed in the introduction. A spring balance and a standard test for identification of positive specimens are equivalent on an abstract level. 

The explicit statement for the spring balance puzzle considered in this article is as such:

\begin{displayquote}
You have N indistinguishable coins and a spring balance. All the coins are indistinguishable without the weighing. 'd' number of coins are counterfeit and and are heavier compared to original one. What is the maximum number of weighing required for the Li's pooling strategy applied in this scenario considering 'g' pools of size 'k'.
\end{displayquote}

 We can compare this with the statement in previous subsection, and as depicted in  Fig. \ref{balancedorfman} (b). If we replace coins by specimens the problems are similar except for the samples in concern and instruments used for the testing. 
 
\section{A quantitative exploration of the Li's S stage algorithm}

We will revise a few insights into Li's S stage procedure \cite{Zhu,Li} for group testing. It is to be noted that Li's s stage algorithm is a generalization of Dorfman's procedure (which has S=2 stages rather than s$>$2). It is also to be noted that we qualitatively showcased the equivalence between the spring balance puzzle and the problem of detection of infection in the human population. Hence, all the quantitative insights of Li's S stage algorithm for infection identification also apply to the spring balance puzzle stated in previous section.

In the Li's S stage algorithm Fig. \ref{Lis}, for a population of size 'N' and 'd' number of positive specimen, the prevalence rate is p=d/N. Consider for 1$^{st}$ stage there are $g_1$ groups of equal size $k_1$ and for i$^{th}$ stage there $g_i$ groups of size $k_i$ (i.e. $g_i=N/k_i$). The number of stages considered are 'S' in number. 

It is to be noted that there can be cases where at most one group can be of size lesser than $k_i$ because the population gets exhausted before filling up that group. We will still consider this group to be of size $k_i$. For example, if there are 9 coins and we group the coins in size of 2, there will be 5 groups in total and there will be one group with only 1 coin. In this case we will still consider this group to be of size 2. Consider $m_1$ groups are found to be having positive specimen in 1st stage and $m_i$ in i$^{th}$ stage.

When there are S=1 stages then the number of querying will be constant i.e. N, irrespective of the prevalence. This is because there will be no grouping at any stage in this procedure. 

When there are S=2 stages (Dorfman's strategy), there will be pooling at 1$^{st}$ stage and individual testing in the second stage. The optimal pool size for first stage in this case will be $k_1=\sqrt{N/d}$. Then the optimal number of groups would be $g_1=\sqrt{Nd}$. For the derivations check \cite{Zhu}. The maximum number of tests required (worst case scenario) is $t=2\sqrt{Nd}$.

For a procedure with S stages, the optimal pool size for i$^{th}$ stage is  $k_i=(N/d)^{(S-i)/S}$ and the optimal number of groups for a particular stage is $g_i \leq d(N/d)^{1/S}$. The maximum number of tests required for all S stages is  $t=Sd(N/d)^{1/S}$.

The above equation only tells us given 'S' stages how much tests would be required in worst case scenario, however we have not fixed optimal number of stages required to reduce the number of tests. Now, the optimal number of stages to exhaust the testing of all the N specimens is $S_0=ln(N/d)=-ln(p)$. Using this knowledge we derive that the maximum number of tests needed for a given population N and d positive specimen when optimal number of stages are implemented, is $t=edln(N/d)$ (obtained by replacing $S_0$), where 'e' is the Euler's constant.

A few things need to be pointed out as they are very important:

1) This pooling strategy may not be a good choice in cases where the prevalence rate '$p$' is above certain threshold. It is to be noted that $p=d/N \leq 1$ as number of positive specimen cannot exceed the population size. If we consider Li's procedure with S=2 stages, then the worst case scenario is  $t=2\sqrt{Nd}=2\sqrt{N^2p}=2N\sqrt{p}$. Until and unless $\sqrt{p}\leq 0.5$ (or $p\leq 0.25$) , $t \leq N$ otherwise it becomes greater than N and pooling is a futile exercise compared to individual testing whose worst case scenario is the constant 'N'. 

Hence, the prevalence rate threshold for $S_0$ stages is when the tests needed are equal to or greater than 'N'. The number of worst case tests required can be written in terms of the prevalence rate, $t=edln(N/d)=eNpln(N/Np)=-eNpln(p)$. Hence when, $t\geq N$ the prevalance rate has to be $-pln(p) \geq 1/e$.

This also has been one of the major hurdles in implementing these strategies when the infection has spread to a larger population. However, we will see in later sections how it can be overcome in some cases inspired by superposition of queries.

\section{Physical implementations of the puzzle}
\subsection{Binary counterfeit coin puzzle}

Let us understand a version of the above problems in a binary form. This would help us in creating logical circuits to demonstrate the problem and understand them in unified manner.

Let the sample population be a binary string made of N binary bits, 0 and 1 (we will call it the secret string 
'$s$'). 0 denoting an original bit  (original coins or negative specimens) and 1 denoting defective bit (counterfeit coin or positive specimens). We are unaware of the nature of the string '$s$', i.e. which particular bit is 1 or 0. We can denote the positions of the bits by an index. For example, in the sample of size 4 and the secret string, s=0011, the 1st bit is $s_1=1$, 2$^{nd}$ bit is $s_2=1$, 3$^{rd}$ bit is $s_3=0$ and 4$^{th}$ is $s_4=0$. 

There exists a query string '$x$' which indicates which particular bit of '$s$' is being queried at a particular step in the solution process. $x_i=1$ denoting $s_i$ being queried in the particular step and $x_i=0$ denoting it being not queried. In the case of the classical spring balance puzzle it is equivalent to i$^{th}$ coin being weighed in or not and in the case of pooled testing, it is equivalent to i$^{th}$ specimen being tested for infection or not. For example, if the query code is 0110, it means the 2$^{nd}$ and 3$^{rd}$ bit of 's' is being queried and others are not. 

The binary version of weighing on a spring balance or testing for infection in specimens is given by the transformation: 
\begin{equation}
f(x)=\sum_{i=1}^{N}x_i s_i=hw(x,s)
\label{Spring Balance Function}
\end{equation}
We  will call it 'binary spring balance' function. The output of the transformation reveals information about the nature of the string $s$ for a particular query $x$.
For the identification of the defective bit, if $f(x)=0$ then there is no defective bit detected by the particular query $x$, otherwise ($f(x)>0$) there is a defective bit identified by the particular query $x$. The summation in the function is nothing but the 'Hamming weight' ($hw(x,s)$) of the bits in the secret string '$s$' which are queried by the particular query string '$x$'. Hamming weight is nothing but the value of the number of 1s in a particular string. This is intuitively similar to difference between the original weight and observed weight on the spring balance scale. Moreover, the value of output also reveals the number of defective bits present, like the spring balance. Note that the value of $f(x)$ is always an integer and is always greater than 0 and lesser than N. 

For example, for the query string x=1011 being used on s=0011 given above: 

$f(x)=1 \times 0 + 0 \times 0 + 1 \times 1 + 1 \times 1=2=10 \text{(in binary)}$ 

Hence, we can conclude that there is at least a defective bit present at index 1, 2 and 4 bit of the string 's'. However, if the a query string x=1100 is applied to the same secret code:

$f(x)=1 \times 0 + 1 \times 0 + 0 \times 1 + 0 \times 1=0 \text{(in binary and decimal)}$ 

Here, no defective bit was found at index 3 and 4. 

As only two binary bits are multiplied at a time, $\times$ can be replaced by an AND gate operation on the two bits. The circuit to implement addition can be done using Full Adder circuits \cite{Mano}.

We now want to identify the defective bits in least number of queries. The procedure is as what we mentioned in previous section using Li's S stage algorithm. We will show by example how it is to be performed in the BSB puzzle case.

Consider a 12 bit secret code 10000000000. There is only one defective bit and hence prevalence rate is p=d/N=1/12. An individual querying (S=1 Li's algorithm) procedure would look like as given below,\\
f(000000000001)=0, \\
f(000000000010)=0, \\
...\\
f(100000000000)=1. \\
The maximum number of queries needed to detect the defective bit would be N=12 for d=1. This was mentioned in the previous section. We can reduce the number of queries by using Li's pooling technique.

The optimal number of stages required, $S_0$ for this problem is around $S_0=ln(12)\approx 2$. Hence, only two stages are needed. We can divide the string in groups of size $k_1=\sqrt{12}\approx 3$, this gives us $g_1=4$. The querying then proceeds by pooling first 3 bits (x=000000000111), then next 3 bits (x=000000111000) and so on. 
\\
f(000000000111)=0,\\
f(000001111000)=0,\\
f(000111000000)=0,\\
f(111000000000)=1.\\

This shows 4 queries are needed in first stage. The defective bit is located in the 4$^{th}$ group.

Now in the 2$^{nd}$ stage, we detect the defective bit by using individual querying as expected by the S=2 procedure of Li's. 
\\
f(001000000000)=0,\\
f(010000000000)=0,\\
f(100000000000)=1.\\

Hence, we see that the last bit (12$^{th}$) in the secret string $s$ is the only defective bit. In general for the BSB puzzle we pool the bits in 1$^{st}$ stage and then query the collective secret string one by one. We detect which groups were having defective bit and then move to next stage. The total number of queries needed for this sample is $t=7$. This value is equal to the worst case queries needed to test a population of size N=12 with prevalence rate of 1/12 ($t=-eNpln(p)\approx 7$).

Thus, we observe that in individual querying for the BSB puzzle we needed 12 queries however using the pooling strategy we required only 7 for the same string size 'N' and prevalence rate 'p'. Hence, pooling reduced the number of queries required in the above example. This can be particularly useful in querying a large secret code with large string size. If the prevalence rate would have been larger than or equal to 0.25, i.e. there were three defective bits, d=3, this procedure would be worse than individual querying. 

\subsection{Creating the circuit for binary spring balance puzzle without superposition of queries}

The circuit is divided in two steps 1) Multiplication of individual bits $x_i$ and $s_i$ and 2) Summation of the output from previous step or finding its' Hamming weight. 

For the first step, as the multiplication is between two bits at a time, it can be implemented using Logical AND gates. In short consider different combination of multiplying 0 and 1 (0x0 = 0, 0x1 = 1, 1x0 = 0 and 1x1 = 1), and compare it with the Logical AND operation (0$\land$0 = 0, 0$\land$1 = 0, 1$\land$0 = 0, 1$\land$1 = 1 ), both provide same output. Hence, the first step can be easily implemented using Logical AND gates.
We create a digital circuit to implement this example.

The next step can be implemented using Full Adder and Half Adder circuits. The minimum number of Full Adder and Half Adder circuits needed to find the hamming weight of an 'N' bit string can be found at \cite{Brandao}. For N=12 bits we need 8 Full Adder and 4 Half Adder circuits.

A simulation of the complete circuit implementation for N=12 bits has been uploaded at \cite{Circuitverse}. To use it the reader is advised to input a particular secret string of size N=12 of their choice. Then the Dorfman's procedure is to be implemented using various queries as mentioned in previous section. The output is the value of the spring balance function for a particular query '$x$'. 
 
 \section{Solution to BSB puzzle using superposition of queries}
The query string in the Li's solution to BSB puzzle uses classical binary bits, asking if a particular bit of the secret string is being queried or not, one at a time: NO=0 and YES=1. However, if a query can exist in a superposition of states of YES and NO (denoted by $\frac{1}{\sqrt{2}}(\ket{0}+\ket{1})$) simultaneously, BV algorithm can be implemented to solve the BSB problem using only single query. Such query problems have been considered by quantum computing experts in the past, however we emphasize that some of these algorithms can also be implemented in classical domain. Superposition states of binary bits are known as 'qubits' in literature. It is to be noted superposition states like qubits can be implemented using classical systems too and in some literature are known as "cebits" \cite{Perez,Arvind,Konrad}. Note that this statement is not in violation of any previous knowledge because, the power of quantum computing lies in superposition and nonlocality combined \cite{Spreeuw,Spreeuw2}. We would caution that different interpretations would have different viewpoints regarding this, however we do not intend to delve into this and rather showcase applicability of superposition of states towards solving the BSB problem at hand. A classical version of a qubit can be easily implemented using polarization of classical light. It is to be noted that the secret string is still binary and do not exist in superposition.

Terhal and Smolin\cite{Terhal} were the first to report connection between counterfeit coin puzzles and the problem addressed by BV algorithm\cite{BV}. In the past Iwama et.al \cite{Iwama} has considered solving the balance puzzle using quantum mechanics. They have aptly named it the "quantum counterfeit coin" puzzle. They however in their article consider a beam balance rather than a spring balance. Due to this they consider a query trit (ternary bit) string rather than a query bit string. A single trit is considered to be of a value either '-1','0' or '1'. -1 representing placing the coin on left pan, 1 representing placing on the right pan and 0 placing it nowhere. Moreover, the Hamming weight of the query string in their case has to be even (i.e. the number of trits in the trit string different than 0 has to be even).

fThe Bernstein-Vazirani\cite{BV} (BV) algorithm was intended to showcase a method to extract a secret string embedded inside a "black box" known as an "Oracle". Simply, the idea of a black box can be considered to be a box holding information of a secret string, which is registered there, maybe by your friend or enemy who does not want to reveal it to you. You catch hold of the box and reveal the secret string from the outputs from this black box by implementing different queries through its' inputs. The motivation is same as the problems considered previously. 

\begin{figure}
    \centering
    \includegraphics[width=0.8\columnwidth]{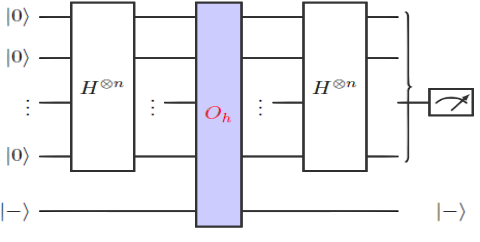}
    \caption{Standard logical circuit diagram for the Bernstein-Vazirani algorithm. The blue box $O_h$ is called oracle. The boxes labelled $H^{\otimes n}$ is a gate with N inputs and N outputs and is an outer product of Hadamard gates. The last symbol represents measurement. Ancilla qubit $\ket{-}$ is depicted at the bottom of the circuit.}
    \label{bv}
\end{figure}

Fig.\ref{bv} is the complete diagram of the circuit implementing BV algorithm. It can be divided into three parts 1) Initialization of query 2) Querying 3) Information retrieval. The blue box in the circuit is a special type of oracle known as phase oracle.The Hadamard gates are denoted by boxes labelled 'H'. Measurements are denoted by the meter box.

Let us look at the algorithm step by step 

1) Initialization of query:\\
  Consider that all states to be used for querying in the algorithm are initialized as $\ket{0}$ which denotes a quantum state. It can be implemented as  horizontally polarized light beam or can be considered to be a spin up electron or a electron in ground state of a two level system. N such states are passed through equal number of Hadamard gates each.

The Hadamard gate transforms as such:
$H\ket{0}=\frac{1}{\sqrt{2}}(\ket{0}+\ket{1})=\ket{+}$ and $H\ket{1}=\frac{1}{\sqrt{2}}(\ket{0}-\ket{1})=\ket{-}$. On the other hand $H\ket{+}=\ket{0}$ and $H\ket{-}=\ket{1}$. These gates in this step create superpositions of $\ket{0}$s and $\ket{1}$s as discussed above. The result of Hadamard gates acting simultaneously on a product state of $\ket{0}$s and $\ket{1}$s of size N denoted by $\ket{a}^{\otimes N}$(for example $\equiv\ket{001..1.0..1}$) is

\begin{equation}
H^{\otimes N}\ket{a}^{\otimes N}=\frac{1}{\sqrt{2^N}}\sum\limits_{x=\{0,1\}^N}(-1)^{\sum\limits_{i=1}^{N}a_ix_i(mod\text{ }2)}\ket{x}^{\otimes N}
\label{Hproduct}
\end{equation}

Here, $\sum\limits_{x=\{0,1\}^N}$ is a sum over all different permutations of 0 and 1 of size N. For example, $x=\{0,1\}^2 \equiv 00,01,10 \text{ and } 11$. Now, if a=01 is some string of size N=2 then 
$H^{\otimes 2}\ket{01}=\frac{1}{\sqrt{2^2}}\ket{+-}=\frac{1}{\sqrt{2^2}}(\ket{00}-\ket{01}+\ket{10}-\ket{11})=\frac{1}{\sqrt{2^2}}\sum\limits_{x=\{0,1\}^2}(-1)^{\sum\limits_{i=1}^2a_ix_i(mod\text{ }2)}\ket{x}$.\\

Hence, when an input $\ket{0}^{\otimes N}\ket{-}$ is transformed by $H^{\otimes N}\otimes I$ then the output is 

\begin{equation}
(H^{\otimes N}\otimes I )\ket{0}^{\otimes N}\ket{-}=\frac{1}{\sqrt{2^N}}\sum\limits_{x=\{0,1\}^N}\ket{x}^{\otimes N}\ket{-}=\ket{+}^{\otimes N}\ket{-}
\label{Hproduct2}
\end{equation}

The R.H.S of above equation is used for querying the oracle in next stage. It is to be noted the final 'N+1'$^{th}$ qubit $\ket{-}$ (called the 'Ancilla") can also be created from $\ket{0}$ by using the X and H gates in series.  For example, the output of $H^{\otimes 3} \otimes I$ acting on $\ket{0}^{\otimes 3} \ket{-}$ is $\ket{+}^{\otimes 3}\ket{-}=\frac{1}{\sqrt{8}}(\ket{000}+\ket{001}+\ket{010}+\ket{100}+\ket{011}+\ket{101}+\ket{110}+\ket{111})\ket{-}$.\\

2) Querying: \\ 
An oracle is a "black box" which takes in several inputs and gives outputs. We usually do not know what it holds and in standard problems try to reveal the hidden property of the constituents of the oracle by querying it.

The standard BV algorithm uses an oracle known as a phase oracle, which encodes a function onto the phase of the inputs.

Hence, when a product state '$\ket{x}^{\otimes N}\otimes \ket{-}$' (x is a binary string), passes through a phase oracle, we have

\begin{equation}
    O_{h}(\frac{1}{\sqrt{2^N}}\sum\limits_{x=\{0,1\}^N}\ket{x}^{\otimes N}\otimes \ket{-})=\frac{1}{\sqrt{2^N}}\sum\limits_{x=\{0,1\}^N}\ket{x}^{\otimes N}\ket{- \bigoplus h(x)} \nonumber
\end{equation}

\begin{equation}
    =\frac{1}{\sqrt{2^N}}\sum\limits_{x=\{0,1\}^N}(-1)^{h(x)}\ket{x}^{\otimes N}\otimes \ket{-}\equiv\ket{-\{d\},+\{d'\}}^{\otimes N}\ket{-}
    \label{phases}
\end{equation}

Note, $\ket{- \bigoplus h(x)}=\frac{1}{\sqrt{2}}(\ket{0 \bigoplus h(x)}+\ket{1 \bigoplus h(x)})$. It can be observed in the second equality that the a phase has been encoded into the input, hence it is known as phase oracle. Here, $h(x)=\sum\limits_{i=1}^{N}s_ix_i (\text{mod 2})$, where s is the secret string which is generally unknown to the one who is querying. Using $\ket{-\{d\},+\{d'\}}\ket{-}$ we introduce a new notation to simplify understanding of this algorithm. Here, $\{d\}$ denotes the set of positive integers which correspond to the indices of those bits in s which are 1, and $\{d'\}$ to those indices which are 0. Then the secret string can be written in a format $s=[1\{d\},0\{d'\}]$. For example, if  $\{d\}\equiv\{1,3\}$ and $\{d'\}\equiv\{2,4\}$ then the secret string is s=0101. $\ket{-\{d\},+\{d'\}}^{\otimes N}$ then denotes a product state of size N, where $\ket{-}$ are at index positions given by $\{d\}$ and $\ket{+}$ at index positions given by $\{d'\}$. 

An interesting observation is that if the phase function $h(x)$ is replaced by the Binary spring balance function $f(x)$ (Eq.\ref{Spring Balance Function}), there will be no change in the output of the oracle. In standard textbooks and articles \cite{BV,Iwama,Terhal,Neilsen}, the oracle is usually considered to be encoding a function $h(x)$ onto the phase. The authors are not aware of any article where $f(x)$ is considered. Quantitatively, $(-1)^{f(x)}= (-1)^{\sum\limits_{i=1}^{N}x_{i}s_{i}}=(-1)^{\sum\limits_{i=1}^{N}x_{i}s_{i}(mod\text{ }2)}=(-1)^{h(x)} $. The function $h(x)$ is not equivalent to the spring balance unlike $f(x)$ as it does not reveal number of defective bits in a particular pool, like a spring balance.

For example, when $\frac{1}{\sqrt{2^3}}\sum\limits_{x=\{0,1\}^3}\ket{x}\ket{-}$ passes through the phase oracle holding a secret string s=101 then, $O_h(\frac{1}{\sqrt{8}}(\ket{000}+\ket{001}+\ket{010}+\ket{100}+\ket{011}+\ket{101}+\ket{110}+\ket{111})\ket{-})=\frac{1}{\sqrt{8}}(\ket{000}-\ket{001}+\ket{010}-\ket{100}-\ket{011}+\ket{101}-\ket{110}+\ket{111})\ket{-}=\ket{-+-}\ket{-}$.\\

3) Information retrieval:\\
Even though the information of the secret string held by the oracle is now encoded in the phase of its' output, the phase however is non-retrievable directly.  We therefore need a step to retrieve it, otherwise the whole test is futile. For this, the Hadamard gates will be once again useful.

If Hadamard gates can create superpositions, they can even break it, i.e. $H\ket{+}=\ket{0}$ and  $H\ket{-}=\ket{1}$. Hence implementing $H^{\otimes N }\otimes I$ on the output from the phase oracle acquired in previous step gives:

\begin{align*}
\frac{H^{\otimes N}\otimes I}{\sqrt{2^N}}\sum\limits_{x=\{0,1\}^N}(-1)^{\sum\limits_{i=1}^{N}x_is_i(mod\text{ }2)}\ket{x}^{\otimes N}\ket{-} = \frac{1}{\sqrt{2^N}^2}\sum\limits_{x,z=\{0,1\}^N}(-1)^{\sum\limits_{i=1}^{N}(x_is_i+x_i z_i)(mod \text{ }2)}\ket{z}^{\otimes N}\ket{-} 
\end{align*}

\begin{equation}
= \frac{2^N}{2^N}(-1)^{2\sum\limits_{i=1}^{N}s_ix_i (mod \text{ }2)}\ket{s}\ket{-}+\frac{1}{2^N}\sum\limits_{x,z=\{1,0\}^N}^{z \neq s}(-1)^{\sum\limits_{i=1}^{N}(s_i+z_i)x_i(mod \text{ }2)}\ket{z}\ket{-}=\ket{s}\ket{-}
\end{equation}

 The first term in the last equality represents the case when z=s, it occurs exactly $2^{N}$ times due to the sum $\sum\limits_{x=\{0,1\}^{N}}$. Moreover, $(-1)^{2\sum\limits_{i=1}^{N}s_ix_i (mod \text{ }2)}$ always has a value equal to 1, because the exponent is always even. Hence, the coefficient of first term is always 1. If the coefficient of that term is 1, then the coefficient of other term has to be equal to zero, due to the normalization criterion in quantum mechanics. Hence, the final output is $\ket{s}\ket{-}$. The secret string can be revealed by measurement in a particular basis of all the states in the product state.

\section{Implementation of the algorithm using polarizers}

We can observe that the state $\ket{-}$ plays no direct role in the computation. It is known as the "ancilla qubit" added at the input to sustain reversibility of the oracle in some circuits. It holds no other significance in the circuit. It is to be noted that for the BV algorithm it can be discarded as it holds no significance for the calculation \cite{Brainis,Du,Arvind,Siewert,Londero}. There are several articles exploring classical implementation of quantum computing algorithms, for example using lenses \cite{Vianna}, using linear optical elements \cite{Arvind} and others \cite{Lemelle,Spreeuw,Sabino}. Noting this, we will discuss a modification of the oracle implementation such that it can be demonstrated using cost effective materials available around. 

\begin{figure}
    \centering
    \includegraphics[width=0.9\columnwidth]{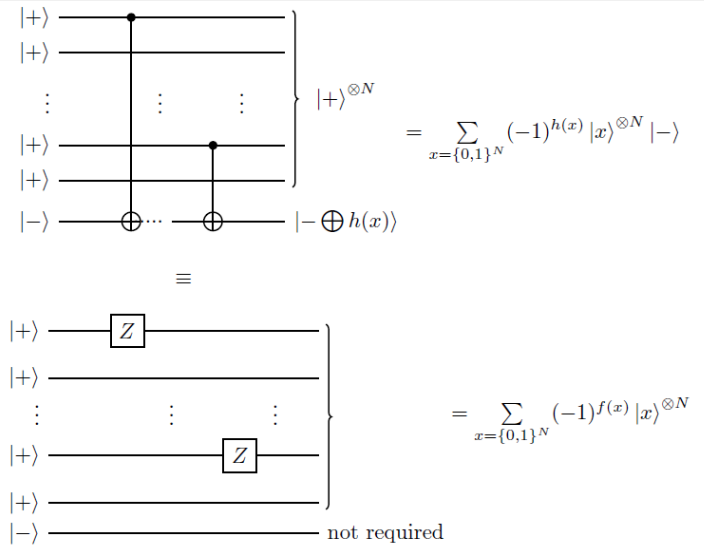}
    \caption{Different Oracle implementations for the same BV algorithm. First using $CNOT$ gates and next using $Z$ and $I$ gates. Note, $f(x)=\sum\limits_{i=1}^{N}x_is_i$ and $h(x)=\sum\limits_{i=1}^{N}x_is_i(mod \text{ }2)$ in the phase does not make any difference.}
    \label{oracle}
\end{figure}

One of the standard ways the phase oracle "$O_{h}$" can be implemented is by using $CNOT$ gates as depicted in Fig. \ref{oracle}.
$CNOT$ gates are two qubit gates and their implementation has been a major problem in optical implementation of quantum algorithms as it requires entanglement \cite{Brien}. Earlier optical implementation of universal quantum computers used non-linear optics \cite{Milburn,Hutchison,Llyod} to implement it. However, later KLM protocol \cite{Knill} solved this issue and this gate could be implemented using linear optical elements \cite{Kok,Brien}. Even, then it is hard to implement this algorithm physically, as it requires some materials which are not readily available. We then ask can we create an equivalent oracle using only single qubit gates like $Z$ and $I$ gates, as the ancilla qubit plays no role in the computation. It has been noted in \cite{Du,Arvind}. However, it is interesting to note that even after this knowledge, an optical implementation was not proposed using only polarizers. All single qubit gates like $Z$ and $I$ can be implemented entirely using only polarizers if polarization is considered for qubit realization \cite{Brien}. First we justify equivalence between two oracle implementations and then show how the latter can be implemented using just polarizers.

The oracle implementation using $CNOT$ gates is as shown in Fig. \ref{oracle}. The diagram depicts $f(x)$ (Eq. \ref{Spring Balance Function}) to be encoded onto the phase rather than $h(x)$ (Eq. \ref{Hproduct}), however computationally it makes no difference. The input states with index position given by the set $\{d\}$ to the oracle act as control to $CNOT$ gates with targets attached to the  ancilla qubit $\ket{-}$ at position 'N+1', depicted by $CNOT[{d},N+1]$. The input states with index position $\{d'\}$ pass through the oracle without any change or they are transformed by $I[{d'}]$. Let us denote this oracle implementation using a notation $CNOT[\{d\},N+1]\otimes I[\{d'\}]$, then

\begin{equation}
(CNOT[\{d\},N+1]\otimes I[\{d'\}])\ket{+}^{\otimes N}\ket{-}=\ket{+}^{\otimes N}\ket{- \bigoplus h(x)}\nonumber
\end{equation}

\begin{equation}
    =\frac{1}{\sqrt{2^N}}\sum\limits_{x=\{0,1\}^N}(-1)^{f(x)}\ket{x}^{\otimes N}\otimes \ket{-}\equiv\ket{-\{d\},+\{d'\}}\ket{-}
    \label{oracleimp1}
\end{equation}

The notation $\ket{-\{d\},+\{d'\}}$ is discussed in previous section.

An implementation of the oracle using the circuit is given by Fig. \ref{oracle} using $Z$ and $I$ gates. It discards the ancilla qubit and still achieves the goal of demonstrating the BV algorithm and solve the Binary spring balance problem. In this implementation the input states with index position $\{d\}$ to the oracle pass through $Z$ gates. The $Z$ gate transforms various states as given: $Z\ket{0}=\ket{0}$ and $Z\ket{1}=-\ket{1}$ and therefore, $Z\ket{+}=\ket{-}$. The input states with index position $\{d'\}$ pass through the circuit without any change. Let us denote this oracle implementation using a notation $Z[\{d\})\otimes I[\{d'\}]$, then

\begin{equation}
(Z[\{d\}\otimes I[\{d'\}])\ket{+}^{\otimes N}=\frac{1}{\sqrt{2^N}}\sum\limits_{x=\{0,1\}^N}(-1)^{f(x)}\ket{x}^{\otimes N}\equiv\ket{-\{d\},+\{d'\}}
   \label{oracleimp2}
\end{equation}

Except for the ancilla qubit $\ket{-}$ the output Eq. (\ref{oracleimp1})  and Eq. (\ref{oracleimp2}) from both oracle implementations are same.

Thus, we have got ridden of the need of 2 qubit gate, $CNOT$ for the solution. The required gates for the demonstration are all single qubit gates, i.e. $H$, $Z$ and $I$ gates, implemented by using just polarizers \cite{Brien,Simon,Bagini}. Hence, only polarizers should suffice to implement our circuit for BV algorithm.

Half wave plates, Quarter wave plates and Linear polarizers are standard optical elements used to manipulate polarization of light. Polarization of incoming light can exist in superposition and can be used for the circuit implementation. According to the Jone's Calculus horizontally polarized light can be considered to be in the state $\ket{0}=(1,0)^T$ and vertically polarized to be in state $\ket{1}=(0,1)^T$, light polarized at 45$^{o}$ is in equal superposition of both polarization $\ket{0}$ and $\ket{1}$ and is denoted by $\ket{+}=\frac{1}{\sqrt{2}}(1,1)^T$ and light polarized at 135$^o$ to be $\ket{-}=\frac{1}{\sqrt{2}}(1,-1)^T$.

Under this nomenclature, Half wave plates (HWP) and Quarter wave plates (QWP)  which are also known as retarders, can be represented in general by the transformation matrix given below:

\begin{equation}
   BF(\eta,\phi,\theta)=e^{-\frac{i \eta}{2}}\left(\begin{array}{cc}
\cos ^{2} \theta+e^{i \eta} \sin ^{2} \theta & \left(1-e^{i \eta}\right) e^{-i \phi} \cos \theta \sin \theta \\
\left(1-e^{i \eta}\right) e^{i \phi} \cos \theta \sin \theta & \sin ^{2} \theta+e^{i \eta} \cos ^{2} \theta
\end{array}\right)
\end{equation}

For Half wave plates,(phase retardation) $\eta=\pi$ and (circularity) $\phi=0$. For Quarter wave plates $\eta=\pi/2$ and $\phi=0$. $\theta$ is the angle between the "fast axis" of the plates and direction of horizontal polarization.

Linear Polarizers (LP) are represented by transformation matrix as given below: 

\begin{equation}
   LP(\theta) = \left(\begin{array}{cc}
\cos ^{2}(\theta) & \cos (\theta) \sin (\theta) \\
\cos (\theta) \sin (\theta) & \sin ^{2}(\theta)
\end{array}\right)
\end{equation}

For demonstration in a lab, these polarizers can be found in form of plastic sheets also known as polymer "retarder films". The linear and quarter wave plates (two quarter wave plates with parallel fast axis in series produce a half wave plate) can also be found in 3D movie glasses using circular polarization for stereoscopy. They can also be made using cellophane tape as suggested in \cite{Velasquez}. 
Let us look at how the gates needed to implement BV algorithm can be created using polarizers.

\begin{equation}
 H\equiv BF(\pi,0,22.5^o)=\frac{e^{\frac{-i\pi}{2}}}{\sqrt{2}}\left(\begin{array}{cc}
1 & 1 \\
1 & -1
\end{array}\right)
\end{equation}

It is transformation matrix for Hadamard gate with an added phase $e^{\frac{-i\pi}{2}}$. The extra global phase $e^{-i\pi/2}$ causes no significant change to the solution intended by our circuit. Optically, if a horizontally polarized light ($\ket{0}$) passes through it, it creates a 45$^{o}$ polarized light ($\ket{+}$), alternatively if a vertically polarized ($\ket{1}$) passes it creates light polarized at 135$^{o}$ $(\ket{-})$, which similar to how Hadamard gate operates. 

The $Z$ gate can be implemented using 

\begin{equation}
    Z \equiv BF(\pi,0,0^o)=e^{\frac{-i\pi}{2}}\left(\begin{array}{cc}
1 & 0 \\
0 & -1
\end{array}\right)
\end{equation}

Optically a $45^o$ linearly polarized light ($\ket{+}$) when passes through such an element it changes to $135^o$ polarized light ($\ket{-}$) with an added global phase which can be ignored. 
\begin{figure}
    \centering
    \includegraphics[width=\columnwidth]{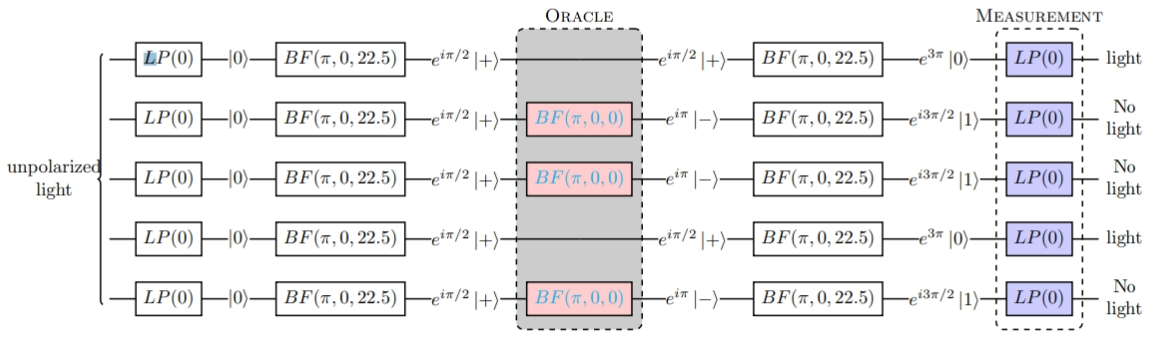}
    \caption{Complete diagram depicting demonstration of BV algorithm using just polarizers. $LP(0^0)$ are linear polarizers with fast axis parallel to the horizontal polarization to be considered. $BF(\pi,0,22.5^o)$ act as Hadamard gates. $BF(\pi,0,0^o)$ are $Z$ gates to be placed at the same index positions where $s_i=1$ (i=1 to N). The bottom most input has index 1.   } 
    \label{opimp}
\end{figure}
Linear polarizers ($LP(0)$) are also needed to convert unpolarized light to horizontally polarized light ($\ket{0}$), for application in the circuit.

\begin{equation}
    LP(0^o)=\left(\begin{array}{cc}
1 & 0 \\
0 & 0
\end{array}\right)
\end{equation}

The procedure for the demonstration is as follows (Fig. \ref{opimp}): 1) Unpolarized light is passed through 'N' Linear polarizers $LP(0^0)$, by switching on a light source. You can consider it as a single query switch turned ON. 2) Then its' output is passed through 'N' number of Half wave polarizer films $BF(\pi,0,22.5^o)$, which act as Hadamard gates. 3) The output from this passes through the oracle which consists of $BF(\pi,0,0^o)$ which act as $Z$ gates. For demonstration, someone can be asked to secretly position  $Z$ gates at desired position, without revealing it to others. 4) The secret string registered in the oracle is now registered in the phase of the output from the oracle, which can be revealed using 'N' Hadamard gates as used before. 
The optical implementation of the BV algorithm circuit using just polarizers and classical light, for a secret code s=10110 registered in the oracle, is as shown in Fig \ref{opimp}.

In terms of the problem of testing for infection in a certain population, if we consider that the positive specimens act as $Z$ gates ($BF(\pi,0,0^o)$) and we use polarization of classical light to query for their presence. The problem can then be solved in just a single query by using the circuit mentioned above. In this case we do not need to even worry about the prevalence rate 'p' of infection in the population as required for the other "classical" pooling strategies described above. We do not even need to care about dilution of samples due to pooling. The limitation of this method however is related to the nature of the specimens. If the query needed to test the specimens cannot exist in superposition for the testing mechanism to work, then it is not possible to test in a single query. Hence, a device and testing mechanism can be created where Hilbert space structure applies, such that a cebit or a qubit can be used for querying. We do not even need to be in quantum domain for implementing such a strategy. In the classical domain we saw Hilbert space structure applies to polarization of classical light and other candidates are Orbital Angular Momeentum of classical light \cite{Singh} or path of the classical light \cite{Anirban} .

\section{Conclusion and discussion}
In this article we presented a unified picture of the Balance puzzles and the problem of pooled testing for infection in a large population. It was showcased that the "Spring balance puzzles" (as opposed to beam balance puzzles) are similar the problem of pooled testing for infection. We then propose a logical version of these problems and name it as Binary spring balance puzzle, not yet discussed earlier in standard articles. It provides a unified logical picture for both the set of problems irrespective of the testing mechanism to be implemented. The logical function $f(x)$ where $x$ is a "binary" query string acts as a spring balance or the device which tests for positive specimen. Circuits are presented to demonstrate the solution to the Binary Spring balance puzzle. First is where the queries are not superimposed like in the case of Li's S stage algorithm and another where the queries are superimposed like in case of the Bernstein-Vazirani algorithm. Even though it is considered that the Bernstein-Vazirani algorithm needs to be implemented on a quantum computer, we show that it can be implemented using classical light and only polarizers which are readily available. This is possible by modification of the Oracle implementation which does not need entanglement. This circuit also demonstrates that in cases where the testing mechanisms allow superposition of queries a device can be created where the testing for infection in a population can be implemented in single query without caring for the prevalence rate or dilution of the samples, which plague other pooling strategies. Bernstein-Vazirani algorithm supersedes all other algorithms which are proposed currently for testing. Moreover, as this demonstration works in classical domain, the device does not need the worry about working in the quantum domain. We hope that this knowledge paves path to development of better testing strategies during pandemics. The demonstration of Bernstein-Vazirani algorithm mentioned in this article is new and surprisingly not discussed elsewhere and it can be easily implemented in an undergraduate lab.

\begin{acknowledgments}

I am grateful to Ayan Biswas for having patience to discuss. 

\end{acknowledgments}

\end{document}